\documentclass[aps,pra,twocolumn,showpacs,preprintnumbers,amsmath,amssymb,citeautoscript,superscriptaddress,bibliography]{revtex4-2}
\usepackage{amsbsy,amssymb,amsmath,bm}
\usepackage{graphicx}
\usepackage{braket}
\usepackage{float}
\usepackage{textcomp}
\usepackage{color}
\usepackage[colorlinks=true,linkcolor=red]{hyperref}
\usepackage[sort&compress]{natbib}
\usepackage[normalem]{ulem}
\usepackage[shortlabels]{enumitem}
\graphicspath{{../Images/}}

\usepackage{float}
\usepackage[caption = false]{subfig}
\usepackage{amsmath}

\input{epsf}

\begin{document}
\title{
Quantum dynamics of disordered arrays of interacting superconducting qubits: signatures of quantum collective states
}
\author{M. V. Fistul}
\affiliation{Theoretische Physik III, Ruhr-Universit\"at Bochum, Bochum 44801, Germany}
\affiliation{National University of Science and Technology ``MISIS", 
Moscow 119049, Russia}

\author{O. Neyenhuys}
\affiliation{Theoretische Physik III, Ruhr-Universit\"at Bochum, Bochum 44801, Germany}

\author{A. B. Bocaz}
\affiliation{Theoretische Physik III, Ruhr-Universit\"at Bochum, Bochum 44801, Germany}

\author{I. M. Eremin}
\affiliation{Theoretische Physik III, Ruhr-Universit\"at Bochum, Bochum 44801, Germany}

\date{\today}

\begin{abstract}
We study theoretically the collective quantum dynamics occurring in various interacting superconducting qubits arrays (SQAs)  in the presence of a spread of individual qubit frequencies. The interaction is provided by mutual inductive coupling between adjacent qubits (short-range Ising interaction) or inductive coupling to a low-dissipative resonator (long-range exchange interaction). In the absence of interaction the Fourier transform of temporal correlation function of the total polarization ($z$-projection of the total spin), i.e. the dynamic susceptibility  $C(\omega)$,  demonstrates a set of sharp small magnitude resonances corresponding to the transitions of individual superconducting qubits. We show that even a weak interaction between qubits can overcome the disorder with a simultaneous formation of the collective excited states. This collective behavior manifests itself by a single large resonance in $C(\omega)$. In the presence of a weak non-resonant microwave photon field in the low-dissipative resonator, the positions of dominant resonances depend on the number of photons, i.e. the \textit{collective ac Stark effect}. 
Coupling of an SQA to the transmission line allows a straightforward experimental access of the collective states in microwave transmission experiments and, at the same time, to employ SQAs as sensitive single-photon detectors.  . 
\end{abstract}

\maketitle


\section{Introduction}
Over the last decade there has been growing interest in various atomic, optical or solid state systems demonstrating the coherent quantum-mechanical dynamics on a macroscopic scale \cite{bruss2019quantum}. Superconducting circuits composed of few small Josephson junctions, i.e. superconducting qubits, appear to be unique solid state systems in which various important factors for the potential applications such as low dissipation and decoherence, electromagnetic field tunable qubits parameters, strong electromagnetic coupling between qubits and a well established measurements setup can be combined 
\cite{bruss2019quantum,krantz2019quantum,kjaergaard2020superconducting,wilhelm2006superconducting}. A large amount of coherent quantum-mechanical phenomena including but not limited to quantum beats, microwave induced Rabi oscillations, Ramsey fringes were observed in experiments with individual superconducting qubits \cite{kjaergaard2020superconducting,wilhelm2006superconducting,peterer2015coherence}. Furthermore, embedding superconducting qubits in a low-dissipative resonator results in a strong coupling between single qubits and cavity photons that has allowed to observe an entanglement between qubits states and Fock photon states \cite{wallraff2004strong,schuster2007resolving}.

A next natural step in this field is to explore the coherent quantum dynamics in low-dissipative \textit{superconducting qubits arrays} (SQAs) composed of a large amount of interacting  superconducting qubits. SQA is an ideal playground to study the collective quantum-mechanical phases and phase transitions in strongly interacting spatially extended quantum systems. For example various SQAs have been used to obtain the ground states of seminal strongly interacting spin systems, namely,  
the spin-glasses \cite{boixo2014evidence}, the $XY$  \cite{king2018observation} and spin-ice models \cite{king2021qubit}. In Ref. \cite{roushan2017spectroscopic} the phenomenon of many-body localization was studied in disordered SQAs composed of nine qubits with nearest-neighbor exchange interaction. A theoretical analysis of coherent quantum dynamics in disordered SQAs was carried out in the absence of interaction \cite{shapiro2015dispersive} and in the presence of Ising nearest-neighbor type interaction \cite{fistul2017quantum}. In the latter case, the collective quantum dynamics has been obtained as the strength of interaction overcomes the disorder in individual qubits frequencies. Experimentally, the collective dynamics was demonstrated spectroscopically in disordered SQAs based on flux \cite{macha2014implementation} and transmon qubits \cite{shulga2017observation,mazhorin2021cavity,brehm2021tunable,brehm2021waveguide}, respectively. In the experimental setups,  Refs.\cite{macha2014implementation,shulga2017observation,mazhorin2021cavity} the SQAs were embedded into a low-dissipative resonator, and it is plausible to assume that the presence of a long-range exchange interaction between well-separated qubits is induced by the excitation (absorption) of virtual photons, i.e. the cavity bus \cite{blais2003tunable,volkov2014collective,majer2007coupling,brehm2021waveguide}.  

In this manuscript we present a detailed theoretical study of collective quantum dynamics in the regime of substantial disorder in qubits frequencies due to unavoidable spread of qubits parameters \cite{mazhorin2021cavity,brehm2021tunable,brehm2021waveguide}. In particular, we consider three different types of interaction between the  qubits: (i) a short-range Ising interaction between adjacent qubits, (ii) artificially introduced global exchange interaction between all qubits, and (iii) a long-range interaction of qubits induced by the coupling to cavity photons. Our analysis is based on the exact diagonalization of the Hamiltonian  and numerical calculation of  the time-dependent correlation function of the total polarization ($z$-projection of the total spin) for SQAs of a moderate size. We analyze the numerical results as a function of SQAs size (number of qubits), an interaction strength and a spread of qubits frequencies (disorder). We show that the collective nature of the quantum dynamics manifests itself by a dominant resonance in the Fourier transform of the time-dependent correlation function, which can be routinely measured in spectroscopic experimental setups with superconducting qubits. 
\cite{macha2014implementation,shulga2017observation,mazhorin2021cavity,bourassa2009ultrastrong}.  Furthermore, for an SQA, coupled to the cavity photon states, the position of a dominant resonance shows a substantial shift depending on the number of photons, i.e. the collective AC Stark effect. This opens a perspective to use SQAs as sensitive single-photon detectors. 

The paper is organized as follows: In Section II we present the model Hamiltonian for  disordered interacting superconducting qubits arrays (SQAs), coupled to low-dissipative resonator and transmission line.  In Section III by making use of the  direct numerical diagonalization of the SQAs Hamiltonian we obtain the dependencies of energy levels and the temporal correlation function of the  total polarization ($z$-projection of the total spin)  of a moderate size SQAs on the interaction strength between the qubits and discuss the collective quantum-mechanical behavior in equilibrium SQAs in the case when the coupling between qubits and a resonator is absent. In Section IV we extend this generic description to a non-equilibrium case when the qubits are strongly coupled to a resonator and obtain the collective ac Stark effect, occurring  in the presence of a  weak electromagnetic field in the resonator. In Section V we propose a complete description of the microwave transmission experiment allowing to observe the coherent collective states. We conclude with Section VI. 

\section{Model of Disordered Interacting SQAs}
\begin{figure}
\centering
\includegraphics[width=3.5in,angle=0]{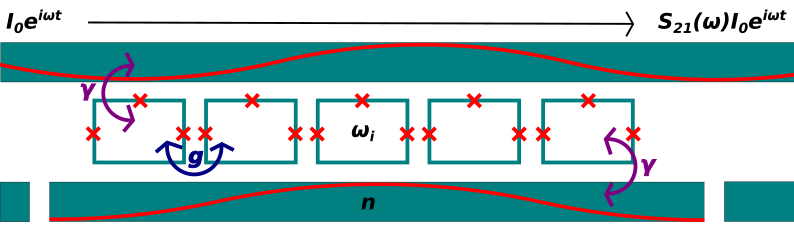}
\caption{ 
Illustrative generic setup of an SQA composed of $N$ flux-qubits. The  SQA is coupled to a low-dissipative resonator (on the bottom) and a transmission line (on the top). Here, $g$ is the strength of mutual inductive coupling between adjacent qubits leading to the Ising (short-range) type of interaction; $\gamma$  is the coupling strength between a single qubit and a resonator; $n$ is a number of microwave photons in a resonator. The  $S_{21}(\omega)$ is the frequency dependent transmission coefficient of electromagnetic waves propagating through the transmission line. The Josephson junctions are marked by crosses.
 }
\label{Pic.1}
\end{figure}

We consider a coherent SQA, composed of $N$ superconducting qubits, {\it e.g.} flux qubits as illustrated in Fig. \ref{Pic.1}. The quantum dynamics of a single qubit is determined by its frequency, $\omega_i$. In a realistic experimental setup the frequencies of individual qubits can differ significantly  due to the presence of an unavoidable spread of physical parameters. An intrinsic short-range coupling $g$ between adjacent qubits is provided by mutual inductance effects. The SQA is also coupled inductively, {\it i.e.} with  current-current interaction $\gamma$, to a low-dissipative resonator (shown on the bottom of Fig.\ref{Pic.1}). Its role is twofold: first, it is used to collect microwave photons of frequency $\omega_0$ at a long time, $Q/\omega_0$, where $Q \gg 1$ is the quality factor of the resonator; second, an inductive coupling between the qubits and the resonator results in an effective long-range interaction between all qubits. The SQA is also coupled to a low-dissipative transmission line (shown on the top of  Fig.\ref{Pic.1}) that is used to experimentally access the coherent quantum dynamics of SQAs and to detect the photon states of the microwave resonator through the measurements of electromagnetic waves transmission coefficient, $S_{21}(\omega)$. 
In order to study the coherent quantum dynamics occurring in a generic setup presented in Fig.\ref{Pic.1} we write the total Hamiltonian as
\begin{equation}
\hat{H}_{tot}= \hat{H}_{SQA}-\alpha I(t)\sum_i \hat \sigma^z_i,
\label{totalHamiltonian}
\end{equation}
where $\alpha$ and $I(t)$ are the coupling strength between the qubits and the transmission line, and the current flowing along the transmission line, respectively. The $\sigma^{x,y,z}$ are the corresponding Pauli matrices. The SQAs Hamiltonian consists of five terms
\begin{multline}
\hat{H}_{SQA}=\hat{H}_{qb}+\hat{H}_{SR}+\hat{H}_{LR}+\hat{H}_{ph}+\hat{H}_{qb-ph}, \  
\label{eq2}
\end{multline}
where the non-interacting part of the qubits Hamiltonian is given by
\begin{equation}
\hat{H}_{qb} = \sum_{i=1}^N \left[\frac{\Delta_i}{2}\sigma^x_i+\frac{\epsilon_i}{2}\sigma^z_i\right]. 
\label{eq3}
\end{equation}
The short-range Ising and the long-range exchange interactions terms  are
\begin{equation}
\hat{H}_{SR} = g_{SR}\sum_{i}\sigma^z_i \sigma^z_{i+1}
\label{eq4}
\end{equation}
\begin{equation}
\hat{H}_{LR} = g_{LR}\sum_{i,j,i \neq j} \left[\sigma^x_i \sigma^x_{j}+\sigma^y_i \sigma^y_{j}\right].
\label{eq5}
\end{equation}
The last two terms in the r.h.s. of Eq. (\ref{eq2}) refer to the photons and their interaction with the qubits 
\begin{equation}
\hat{H}_{ph} = \hbar \omega_0 \hat{a}^{\dag}\hat{a}
\label{eq6}
\end{equation}
\begin{equation}
\hat{H}_{qb-ph} = \gamma \sum_{i=1}^N \sigma^z_i ( \hat{a}^{\dag}+\hat{a}).
\label{eq7}
\end{equation}
Here, $\hat a^{\dag} (\hat a) $ are the creation (annihilation) photon operators. As the wavelength of electromagnetic waves in both the resonator and the transmission line is much larger than the SQA size, we assume that the coupling strengths $\alpha$ and $\gamma$ do not vary along the SQA. From Eq. (\ref{eq3}) one obtains the individual qubits frequencies as $\omega_i=\sqrt{\Delta_i^2+\epsilon_i^2}/\hbar$, where the qubits parameters $\Delta_i$ and $\epsilon_i$ can be varied in experiment by externally applied magnetic fields. In disordered SQAs a substantial spread of individual qubits frequencies takes place \cite{mazhorin2021cavity,brehm2021tunable,brehm2021waveguide}.
A spread of qubits frequencies, i.e a strength of disorder, will be quantitatively characterized by relative variations of qubits frequencies, $\sigma=\sqrt{<(\omega_{i}-\bar{\omega})^2>}/\bar{\omega}$, where $\bar{\omega}=<\omega_i>$ is the average qubits frequency in the SQA and $<...>$ means the averaging over disorder in qubits frequencies. We study an interplay between the disorder and the interaction strength in a most relevant regime for experiments, $\bar{\omega}>g/\hbar$.

To conclude this section we notice that the Hamiltonian, $\hat{H}_{tot}$, is quite general and describes the coherent quantum dynamics in SQAs, composed of other types of superconducting qubits, e.g. charge qubits or transmons \cite{krantz2019quantum,wallraff2004strong,brehm2021waveguide}.

\begin{figure}
\centering
\subfloat[a)]{\includegraphics[width=3.4in,angle=0]{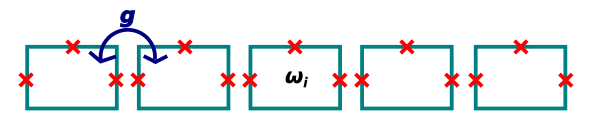}}\\
\subfloat[b)]{\includegraphics[width=3.5in,angle=0]{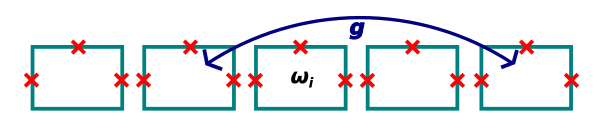}}\\
\subfloat[c)]{\includegraphics[width=3.5in,angle=0]{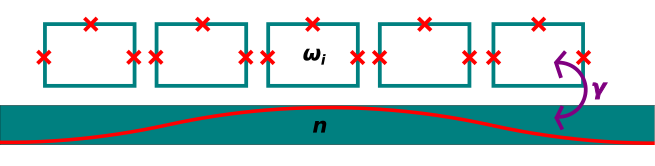}}
\caption{ 
(a) Schematic representation of a disordered SQN with short-range interaction between adjacent qubits (a), an artificial long-range exchange interaction between all qubits (b), or coupled to a low-dissipative resonator (c). $n$ refers to the microwave photon state. }
\label{Pic.2}
\end{figure}

\section{Coherent quantum dynamics in disordered interacting SQAs: equilibrium case}
In what follows, we study the coherent quantum dynamics in various SQAs by a  direct numerical diagonalization of the Hamiltonian, Eq.(\ref{eq2}), for various SQAs sizes ($N=2 \div 8$) and  obtain the eigenvalues $E_n$ and the eigenvectors $\Psi_n$ of a whole SQA. In this section we apply our numerical analysis to the \textit{equilibrium} SQAs assuming the coupling between qubits and a resonator is absent (see, Fig. \ref{Pic.2}(a),(b)). Upon obtaining the $E_n$ and $\Psi_n$ we construct  an explicit expression for the equilibrium temporal correlation function of a total polarization ($z$-component of total spin), $C(t)$,
\begin{multline}
C(t)=\frac{1}{Z}\left\langle \left[\sum_{i=1}^N\hat{\sigma}^z_i(t)\right]\left[\sum_{i=1}^N\hat{\sigma}^z_i(0)\right] \right\rangle= \\
=\frac{1}{Z}\sum_{m,n} e^{-i(E_m-E_n)t/\hbar}e^{-E_n/(k_B T)} \left| \left\langle \Psi_m \left|\sum_{i=1}^N\hat{\sigma}^z_i\right| \Psi_n \right\rangle \right|^2,
\label{TDCF}
\end{multline}
where $Z=\sum_n \exp[-E_n/(k_B T)]$ is the partition function of an SQA.
Note that at low temperatures, $k_B T \ll \min\{E_n,E_m \}$, the main contribution to the summation over $n$ in Eq. (\ref{TDCF}) comes from the ground state $\Psi_0$ with the corresponding eigenvalue $E_0$, only. 
It is convenient to characterize the coherent quantum dynamics of SQAs by the dynamic susceptibility $C(\omega)$:
\begin{equation}
C(\omega)=\lim_{t_0 \rightarrow \infty} \frac{1}{t_0}\int_0^{t_0} dt ~e^{i\omega t}\Im m C(t)
\label{DynSus}
\end{equation}
\subsection{Collective quantum states in disordered SQAs with a short-range Ising interaction}
We start the analysis by looking on the coherent quantum dynamics of disordered linear SQAs in the presence of a \textit{short-range Ising interaction } between adjacent qubits only, see Fig. \ref{Pic.2}(a). 
In this case, Eq.(\ref{eq2}) reduces to the following Hamiltonian 
\begin{equation}
\hat{H}_\textnormal{SQA-SR} =\sum_{i=1}^N \left[\frac{\Delta_i}{2}\hat{\sigma}^{x}_i+\frac{\epsilon}{2}\hat{\sigma}^{z}_i+g \hat{\sigma}^{z}_i \hat{\sigma}^{z}_{i+1} \right]
\label{SQA-SR}
\end{equation}
For the sake of simplicity, we assume all $\epsilon_i=0$. 
The coherent quantum dynamics of such SQAs is determined by relative values of interaction $g$ and disorder $\sigma$ strengths. In our numerical simulations we choose the ranges of these parameters as  
$|g|/\hbar <0.3 \bar{\omega}$ and $\sigma < 0.2$.

\begin{widetext}

\begin{figure}
\centering
\includegraphics[width=\linewidth,angle=0]{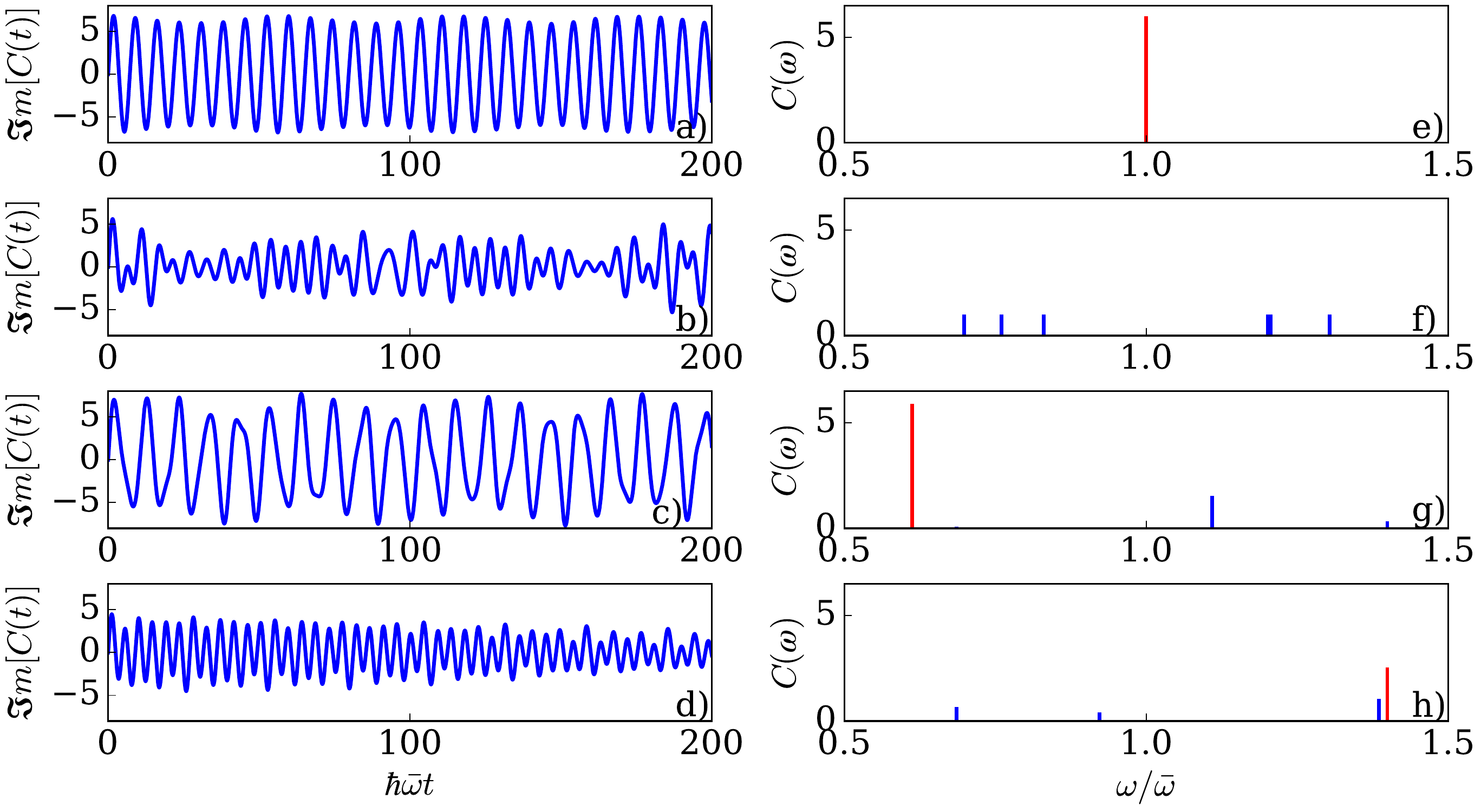}
\caption{ 
The results for the time-dependent correlation function of the $z$-projection of the total spin, $\Im m C(t)$, ((a)-(d)) and the corresponding dynamic susceptibility, $C(\omega)$, ((e)-(h)), which demonstrate a set of resonances for various SQAs. Here we choose  $\sigma=0$ and $g=0$ (a),(e); $\sigma=0.2$ and $g=0$ (b),(f); $\sigma =0.2$ and $g/\hbar= -0.2 \bar{\omega}$  (c),(g); and $\sigma =0.2$ and $g/\hbar= 0.2\bar{\omega}$ (d),(h). The dominant resonances (spikes) are shown by red lines. The total number of qubits is $N=6$. 
}
\label{Pic.3}
\end{figure}

\end{widetext}

The typical dependencies of $\Im m C(t)$ and corresponding $C(\omega)$ for the SQAs with $N=6$ interacting qubits are shown in Fig. \ref{Pic.3}. In the absence of both the disorder and the interaction, the  $\Im m C(t) $ demonstrates precise oscillations at a single frequency, $\omega_{qb}=\Delta/\hbar$ , see Fig.\ref{Pic.3}(a). In this case the dependence of $C(\omega)$ demonstrates a single resonance (spike) (Fig.\ref{Pic.3}(e)). In the presence of disorder for zero interaction strength,  $g=0$, the $\Im m C(t)$ displays a set of oscillations with different frequencies and a set of equal small amplitude resonances in $C(\omega)$, as shown in Fig. \ref{Pic.3}(b),(f). However, for moderate strength of the interaction, e.g. $|g|/\hbar=0.2\bar{\omega}$, the oscillations with all frequencies except a single one are diminished, and one observes the dominant resonance in $C(\omega)$ as shown in Figs. \ref{Pic.3}(c),(d) and \ref{Pic.3}(g),(h). The substantial asymmetry in the $\Im m C(t)$ ($C(\omega))$ for positive (negative) values of $g$ was also observed. Notice that the effects of low dissipation and decoherence would lead to a tiny broadening of the resonances (spikes). 


\begin{widetext}

\begin{figure}[t]
\includegraphics[width=\linewidth,angle=0]{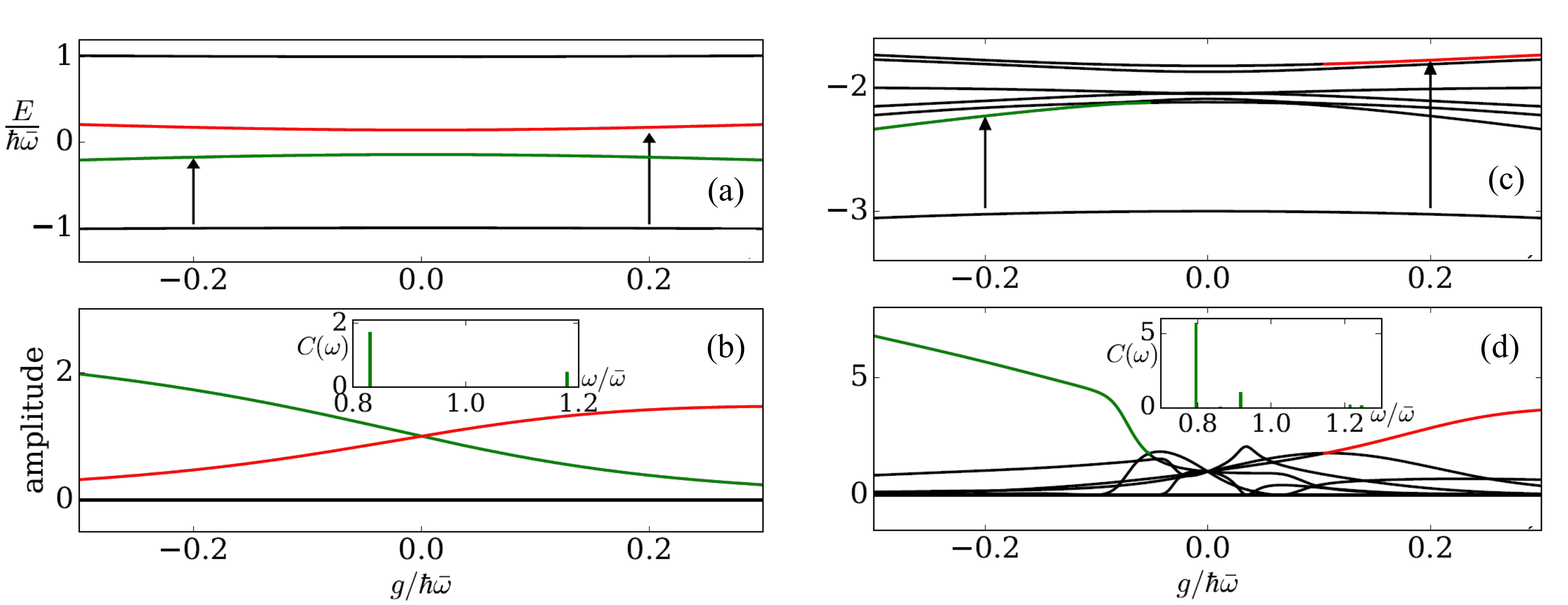}
\caption{ 
Calculated dependencies of energy levels (a),(c) and the amplitudes of  resonances in $C(\omega)$ (b),(d) on the interaction strength, $g$ in the presence of a substantial spread of qubit frequencies $\sigma =0.2$ for the number of qubits $N=2$ (left panels) and $N=6$ (right panels). The photon-induced transitions resulting in the dominant resonances are indicated by arrows. The insets shows the $C(\omega)$ dependence for $g/\hbar= -0.2 \bar{\omega}$. 
}
\label{Pic.5}
\end{figure}

\end{widetext}

To obtain deeper insight on the quantum dynamics of SQAs in the presence of both interaction and disorder we systematically study the dependence of energy spectrum $E_n$ and the amplitudes of resonances $A(g)$ in $C(\omega)$ on the interaction strength $g$. The results for two different SQAs with $N=2$ and $N=6$ are presented in Fig. \ref{Pic.5}. In the limit of $g/\hbar \ll \bar \omega$ one finds a tiny decrease of the SQAs ground state energy upon varying $|g|$. The ground state is well separated from the group of $N$ low-lying excited levels forming the so-called lowest band as shown in Fig. \ref{Pic.5}(a),(c). Consequently, $C(\omega)$ demonstrates a set of $N$ resonances (spikes) whose magnitudes are determined by the interaction strength as evident from Fig. \ref{Pic.5}(b),(d).  The physical origin of these resonances is the photon-induced transitions between the ground state and excited levels of a system. Most of these resonances diminish upon increase of $|g|$ while one of them is strongly enhanced. This dominant resonance indicating the presence of \textit{collective} quantum excited state is determined by a particular transition between the ground state and the lower (upper) excited energy levels for negative (positive) values of the interaction strength $g$. These transitions are indicated by arrows in Figs. \ref{Pic.5}(a),(c). Correspondingly, the frequency position of the dominant resonance is shifted to larger (smaller) values for positive (negative) values of $g$. 

\begin{figure}
\includegraphics[width=\linewidth,angle=0]{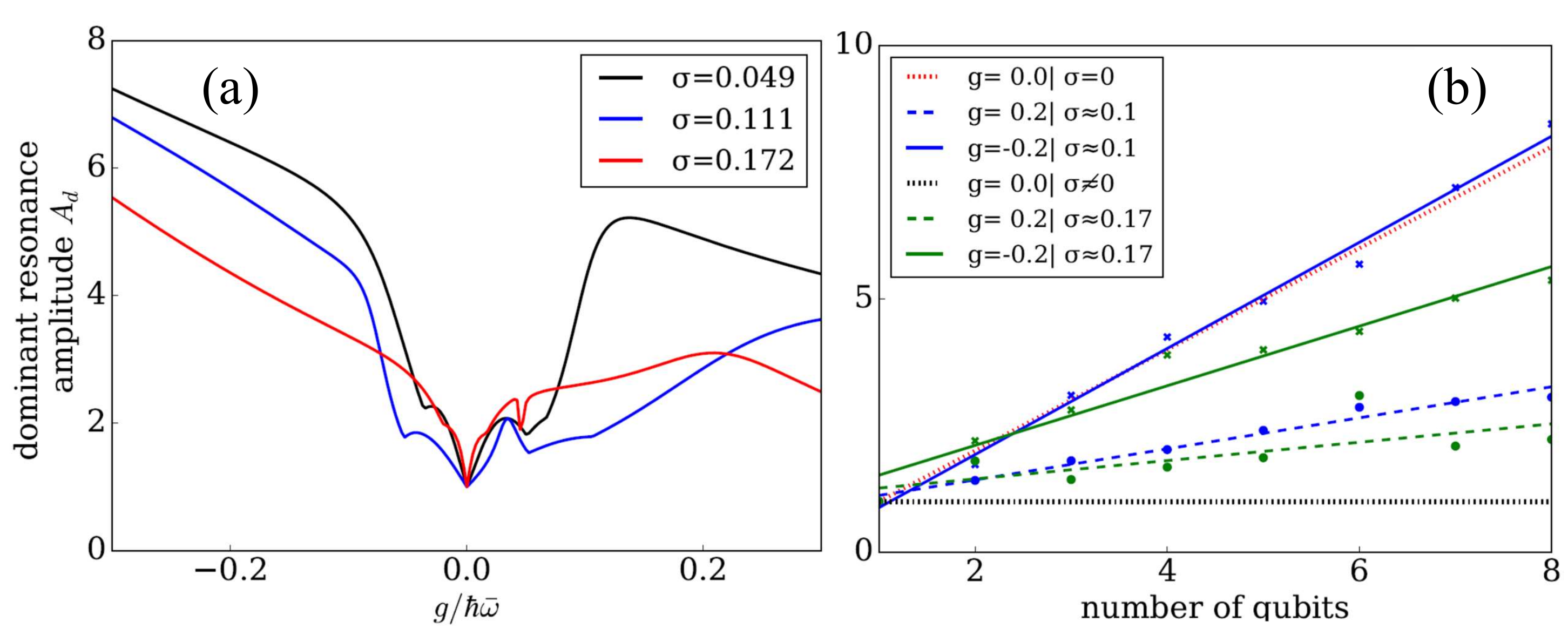}
\caption{ 
Calculated dependencies of the magnitude of dominant resonance $A_d(g)$ in $C(\omega)$  on the interaction strength (a) for 6 qubits, and  on the number of qubits (b) for various values of disorder $\sigma$ and interaction strength, respectively.  
}
\label{Pic.8}
\end{figure}

We also numerically study the dependence of the dominant resonance magnitude, $A_d(g)$, on the strength of disorder, i.e. $ \sigma$. The results presented in Fig. \ref{Pic.8} (a) clearly demonstrate that the $A_d$ increases drastically as the interaction strength overcomes the disorder spread, i.e. $g/\hbar > \sigma \bar \omega$. Correspondingly, studying the SQAs with different numbers of qubits $N$ and fixing the  values of disorder and strength of interaction we obtain that the magnitude of dominant resonance $A_d$ linearly grows with $N$ (see, red and blue lines in Fig.\ref{Pic.8}(b)).

\subsection{Collective quantum states in disordered SQAs with an artificial long-range exchange interaction}

In this subsection we study the coherent quantum dynamics of disordered linear SQAs with an artificial \textit{long-range exchange interaction} between qubits, see Fig. \ref{Pic.2}(b). 
We consider a particular type of long-range interactions, i.e. so-called the global interaction where each qubit interacts with all other qubits with the strength $g$, and  in this case, Eq.(\ref{eq2}) reduces to the following Hamiltonian, i.e. Eqs. (\ref{eq3}) and (\ref{eq5}),
\begin{equation}
\hat{H}_\textnormal{SQA-LR} =\sum_{i=1}^N \left[\frac{\Delta}{2}\hat{\sigma}^{x}_i+\frac{\epsilon_i}{2}\hat{\sigma}^{z}_i\right]+ g\sum_{i,j, i \neq j} \left[\sigma^x_i \sigma^x_{j}+\sigma^y_i \sigma^y_{j}\right].
\label{SQA-LR}
\end{equation}
For the sake of simplicity we assume here that the disorder occurs in the diagonal qubits parameters $\epsilon_i$ only, and the off-diagonal parameter $\Delta$ does not fluctuate. We explore the ranges of the parameters of Hamiltonian, Eq. (\ref{SQA-LR}), $|g|/\hbar <0.2 \bar{\omega}$ and $\sigma  \simeq 0.1$. 

\begin{widetext}

\begin{figure}
\centering
\includegraphics[width=\linewidth,angle=0]{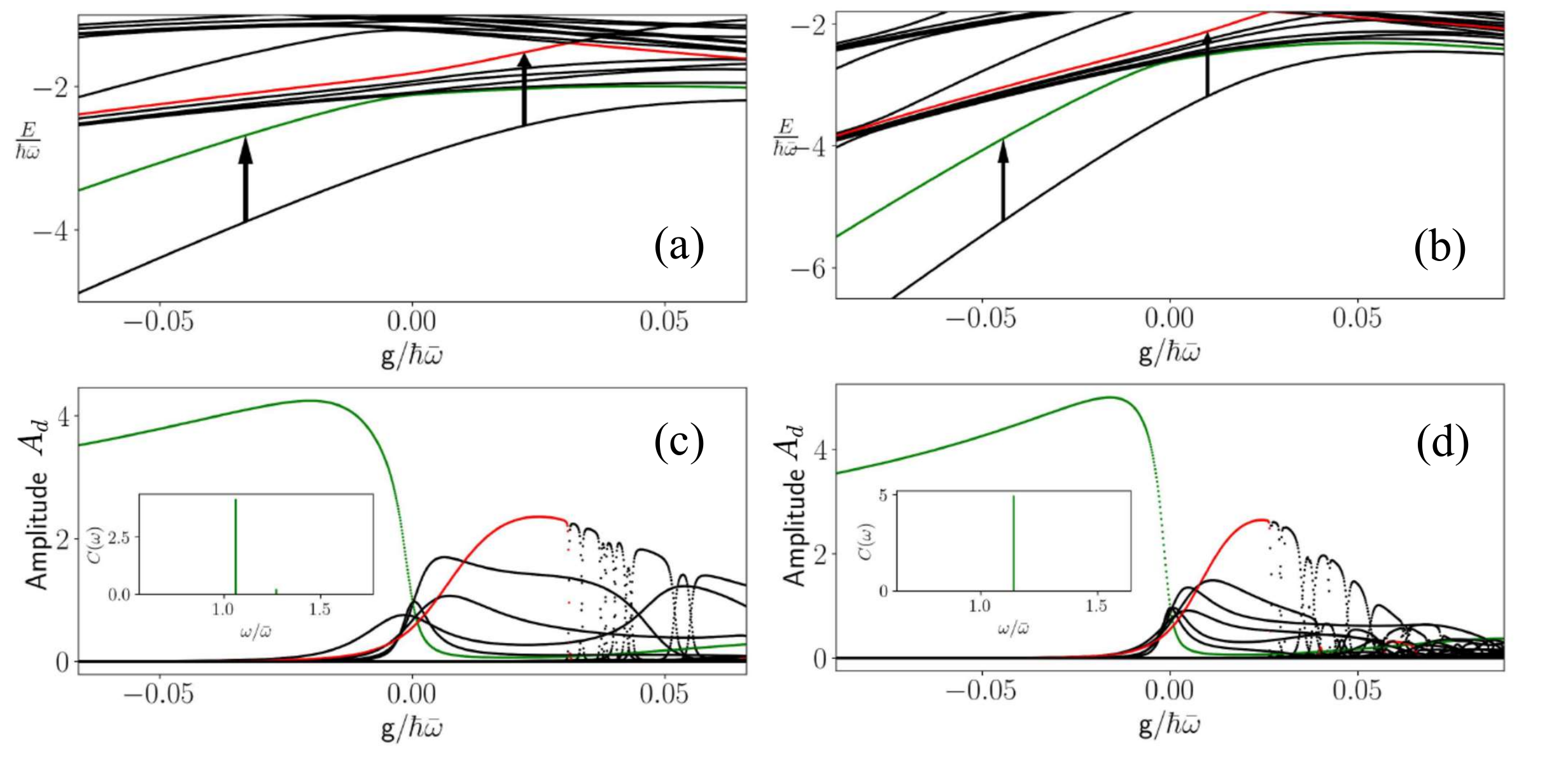}
\caption{ 
Calculated dependencies of low-lying energy levels (a),(b) and amplitudes of resonances (c),(d) on the interaction strength $g$ for the disordered SQN composed of $N=6$ (left panels) and $N=7$ (right panels) qubits. The resonant transitions that can be observed in the microwave transmission experiment are shown by arrows. The insets show the frequency position of the dominant resonance (spike) for $g=-0.033 \hbar \bar{\omega}$ (left panel) and $g=-0.044 \hbar \bar{\omega}$ (right panel). The disorder strength $\sigma \simeq 0.12$ was chosen.
}
\label{Pic.10}
\end{figure}

\end{widetext}

Our typical results are presented in Fig. \ref{Pic.10} for $N=6$ (left panels) and $N=7$ (right panels) qubits, where we show the dependencies of the low-lying energy levels $E_n$ (a),(b)  and the amplitudes of various resonances (spikes) in the dynamic susceptibility $C(\omega)$ (c),(d) on the interaction strength $g$. 
Similarly to the SQAs with a short-range Ising interaction we observe a single large amplitude dominant resonance as the strength of interaction $g$ increases. The dominant resonances indicating the presence of \textit{collective} quantum excited state are determined by the particular transition between the ground state and the lower excited energy levels for negative (positive) values of the interaction strength $g$. These transitions are indicated by arrows. Correspondingly, the frequency position of the dominant resonance is shifted to larger (smaller) values for negative (positive) values of $g$ (shown in the insets of Figs. \ref{Pic.10} for negative values of $g$). 

Comparing the panels in the Figs. \ref{Pic.10} one observes an increase of the magnitude of the dominant resonance with a number of qubits $N$. Also we notice a substantial asymmetry in the dependence $A_d(g)$ on $g$ that is due to the disorder in diagonal qubits parameters $\epsilon_i$ but not in the off-diagonal $\Delta_i$. However, at variance with the short-range Ising interaction in the presence of global exchange interaction  the dominant resonance and ,therefore, the collective excited states form for much smaller values of the interaction strength $g$, i.e. $g/(\hbar \bar \omega) \ll \sigma$. 

\section{Coherent quantum dynamics in disordered interacting SQAs: non-equilibrium case}

\begin{widetext}

\begin{figure}
\centering
\includegraphics[width=\linewidth,angle=0]{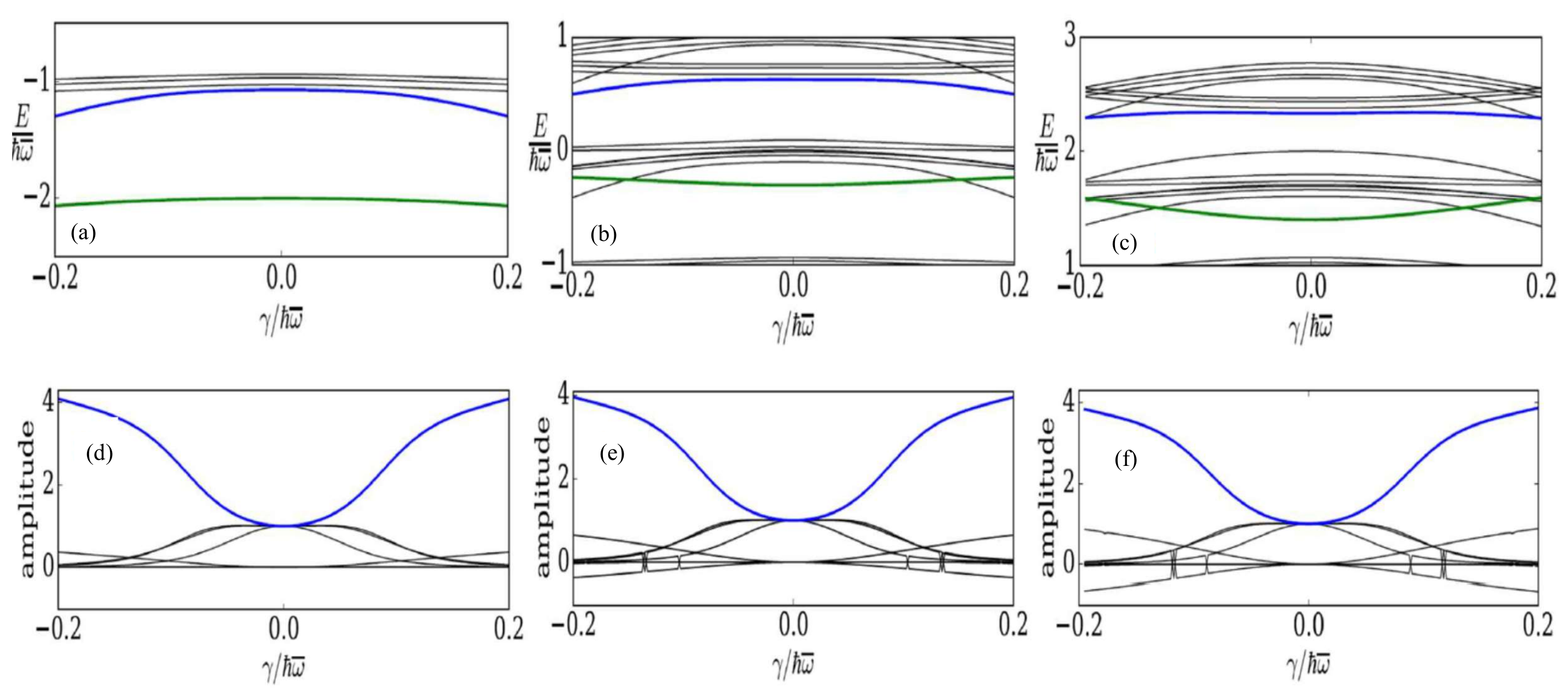}
\caption{ 
The dependence of low-lying energy levels (a)-(c) and the amplitudes of resonances (d)-(f) on the coupling strength $\gamma$ for the disordered SQAs composed of 4 qubits and 4 photon states for $C_0(\omega)$ (left panels),  $C_1(\omega)$ (middle panels), and $C_2(\omega)$ (right panels). The dominant resonances correspond to the transitions between energy levels indicated by green and blue lines in (a)-(c). The disorder strength is fixed as $\sigma=0.1$.  
}
\label{Pic.11}
\end{figure}

\end{widetext}

Next, we turn to a  study of the coherent quantum dynamics of disordered SQAs coupled to a low-diisipative resonator. This setup is schematically shown in Fig. \ref{Pic.2}(c).  To do so we perform a direct numerical diagonalization of the Hamiltonian
\begin{equation}
\hat{H}_{SQA}=\hat{H}_{qb}+\hat{H}_{ph}+\hat{H}_{qb-ph}, \  
\label{Res-qubit-Ham}
\end{equation}
where the Hamiltonians $\hat{H}_{qb}$, $\hat{H}_{ph}$, and  $\hat{H}_{qb-ph}$ are determined by Eqs. (\ref{eq3}), (\ref{eq6})  and (\ref{eq7}), respectively. Here we set the resonant frequency $\omega_0=1.3 \bar \omega$, and the relative variation of qubits frequencies, $\sigma =0.1$. Similarly to the Sec. IIIA we put all values of $\epsilon_i=0$.  The total Hilbert space of the Hamiltonian $\hat{H}_{SQA}$ is the product of $2^N$ qubits states and $n$ photons Fock states. In our numerical analysis the size of SQAs was varied from $N=2$ to $N=6$, and we truncate the number of Fock states to $n=4$. We verify that an increase of Fock states number does not change the coherent quantum dynamics in which the photon states with $|0>$ (zero oscillations, no photons), $|1>$ (a single photon) and $|2>$ (two photons) are involved. Our generic strategy is to fix the coupling strength $\gamma$ and obtain numerically the low-lying eigenvalues $E_i$ and their eigenvectors $\Psi_i$ of the total system. This allows then to calculate the non-equilibrium temporal correlation function $C(t;n)$ for different values of $n$ as 
\begin{multline}
C_n(t)=\left\langle \Psi_i (n)  \left| \left[\sum_{i=1}^N\hat{\sigma}^z_i(t)\right]\left[\sum_{i=1}^N\hat{\sigma}^z_i(0)\right] \right| \Psi_i (n)  \right\rangle.
\label{NEQ-TDCF}
\end{multline}
For our numerical calculations we choose the $\Psi_i (n)$ as the eigenstate of the Hamiltonian (\ref{Res-qubit-Ham}) \textit{maximally overlapped } with the state  $|\downarrow \downarrow...\downarrow \otimes n \rangle $, where $|\downarrow \downarrow...\downarrow \rangle$ is the ground state of noninteracting SQA and $|n>$ is the particular Fock state trapped in the resonator. The non-equilibrium photon state dependent dynamic susceptibility $C_n(\omega)$ is determined by equation analogous to (\ref{DynSus}). After that we vary the parameter $\gamma$ and obtain the dependencies of $E_i$ and $C_n(\omega)$ on $\gamma$. The results are presented in Fig.\ref{Pic.11} for the disordered SQA composed of $N=4$ qubits and different Fock states. 

Similarly to equilibrium disordered SQAs studied in Sec. III, the frequency dependent dynamic susceptibility $C_n(\omega)$ contains a large set of resonances, and the amplitude of a single resonance strongly increases with the coupling strength, $|\gamma|$, see Fig. \ref{Pic.11}(d)-(f). Such resonances with an enhanced amplitude indicate the presence of a strong effective interaction between qubits in SQAs allowing to overcome the spread of individual qubits frequencies, and establish the collective dynamics in disordered SQAs. The effective interaction  arising in SQAs strongly coupled to a low-diisipative cavity is due to an exchange (absorption and emission)  of virtual photons between well separated qubits \cite{blais2003tunable,volkov2014collective,majer2007coupling,brehm2021waveguide}.   

\begin{figure}
\includegraphics[width=\linewidth,angle=0]{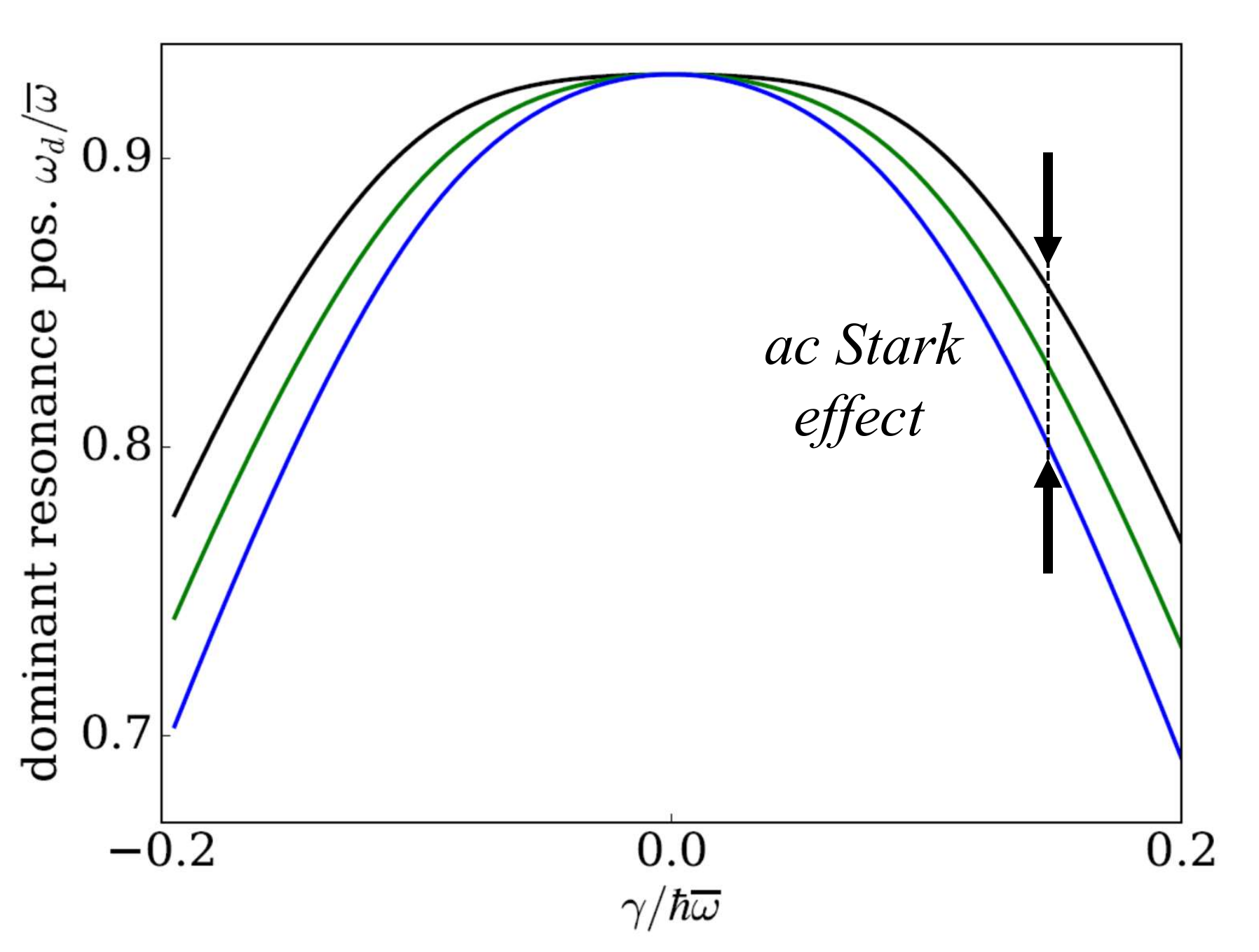}
\caption{ The dominant resonance frequency positions as a function of the interaction strength for different Fock states trapped in the cavity: $|0>$-black line; $|1>$-green line; $|2>$-blue line.
The disorder strength is fixed as $\sigma=0.1$. 
The arrows and the dashed line indicate the {\it ac Stark effect} for a given value of $\gamma$.}
\label{Pic.6}
\end{figure}

The frequency positions of dominant resonances in $C_n(\omega)$ are determined by the specific photon-induced transitions with conservation of the photon number state $|n>$. The energy levels involved in these transitions are indicated by colour in Fig. \ref{Pic.11}(a)-(c). Comparing the energy levels differences we obtain  the characteristic shift of the dominant resonance frequency depending on the number of photons trapped in the cavity, $|n>$.  (see, Fig. \ref{Pic.6}). This effect is nothing else than the \textit{the collective AC Stark effect} and is a central result of our study.

Note that the collective  AC Stark effect in disordered SQAs coupled to a low-dissipative resonator can be qualitatively explained as follows. Exchange of virtual photons results in a strong exchange interaction between well separated qubits, and a system is described by Hamiltonian 
\begin{multline}
\hat{H}_{SQA-res}=\sum_{i=1}^N \left[\frac{\Delta_i}{2}\sigma^x_i+\frac{\epsilon_i}{2}\sigma^z_i\right]+ \hat U_{int}\{\hat{\vec{\sigma}}_1,...\hat{\vec{\sigma}}_N\}+ \\
+\hbar \omega_0 \hat{a}^{\dag}\hat{a}+\gamma \sum_{i=1}^N \hat{\sigma}_i (\hat{a}^{\dag}+\hat{a}),~~~~~~~~~~
\label{SQA-res}
\end{multline}
where $\hat U_{int}$ is the effective interaction between qubits.
As the interaction overcomes the disorder the first three terms in the r.h.s of Eq.(\ref{SQA-res}) establish the collective quantum dynamics described by Hamiltonian $\hat H^{eff}_{0}=\frac{\hbar \omega^{(0)}_d(\gamma)}{2}\hat \sigma^z$, where $\omega^{(0)}_d$ is the frequency of the dominant resonance without taking into account the AC Stark effect. The quantum-mechanical averaging of the last term in the r.h.s of Eq.(\ref{SQA-res}) is obtained in a second order perturbation theory
\begin{multline}
<\gamma \sum_{i=1}^N \sigma^z_i (\hat{a}^{\dag}+\hat{a})> = \\
=<\gamma^2 (\hat{a}^{\dag}(t)+\hat{a}(t))\int_0^t ds \Im m C(t-s)(\hat{a}^{\dag}(s)+\hat{a}(s)) >, 
\label{averaging}
\end{multline}
and the time dependent correlation function $C(t)$ is written as $\Im m C(t-s)=\frac{i}{\hbar}< [\sum_{i} \sigma^z_i(t), \sum_{i} \sigma^z_i(s)]>$. Here, the averaging is assumed over the $n$-th photons Fock state, and the ground (exciting) collective states of an SQA. It results in the effective Hamiltonian 
\begin{equation}
H^{eff}= H^{eff}_{0}+\frac{\gamma^2 <\hat{a}^{\dag}\hat{a}+1/2>A_d(\gamma)}{\hbar (\omega^{(0)}_d-\omega)}\hat \sigma_z
\label{Eff-Hamiltonian}
\end{equation}
and, therefore, the frequency position of dominant resonances demonstrates the collective AC stark shift as $\omega_d(n,\gamma)=\omega^{(0)}_d+\frac{\gamma^2 (n+1/2) A_d(\gamma)}{\hbar (\omega_d-\omega)}$, which is in agreement with our numerical calculations.

To conclude this Section we further notice that as the mixture of photon states are present, e.g. the coherent state, the thermal state etc., the dynamic susceptibility of disordered SQAs,  $C_{ph}(\omega)$, can be obtained through the following expression:
\begin{equation}
C_{ph}(\omega)=\sum_n P(n)C_n(\omega),
\label{Photon-Suscept}
\end{equation}
where $P(n)$ is the probability to obtain the photon in the Fock state $|n>$. 

\section{Spectroscopic measurements of coherent quantum dynamics in SQAs}

The dynamic susceptibility $C(\omega)$ ($C_{ph}(\omega)$) can be directly measured in so-called transmission measurement setup as the studied quantum system is weakly coupled to a low-dissipative transmission line (see the schematic in Fig. \ref{Pic.1}). Indeed,
in the presence of an SQA the transmission coefficient $S_{21}$ is suppressed, and such suppression $\Delta S_{21}(\omega)$ is determined as
\begin{equation}
\Delta S_{21} \simeq \frac{\langle \hat M=\sum_i \hat \sigma^z_i \rangle}{I_0 e^{i\omega t}},
\label{transmission}
\end{equation}
where a probe microwave signal $I_0 e^{i\omega t}$ is applied, and the averaging $\langle...\rangle$ takes place over quantum-mechanical states of the Hamiltonian (\ref{totalHamiltonian}). By making use of a standard response theory analysis \cite{dittrich1998quantum} we obtain the $\Delta S_{21}(\omega)$ as following:
\begin{equation}
\Delta S_{21}(\omega) \simeq \frac{i}{\hbar} \int_0^t d\tau \left \langle [\hat M(t-\tau),\hat M(0)] \right \rangle  exp(i\omega \tau),
\label{transmission-2}
\end{equation}
where
\begin{equation}
\hat M (t)=\exp \left [\frac{it}{\hbar}\hat H_{SQA} \right ]\sum_i \hat \sigma^z_i \exp \left [-\frac{it}{\hbar}\hat H_{SQA} \right ].
\label{operator-M}
\end{equation}
and one finally obtains 
\begin{equation}
\Delta S_{21}(\omega) \simeq C(\omega)
\label{Final-transmission}
\end{equation}
This indicates that the observed AC-Stark effect can be directly measured in the available experimental setups.

\section{Conclusions}
In conclusion, we have studied in detail the coherent quantum dynamics of SQAs composed of $N=2 \div 8$  interacting qubits in the presence of substantial spatial disorder. Our analysis was based on an exact diagonalization of the many-body SQAs Hamiltonian (\ref{eq2})-(\ref{eq7}), and  a subsequent computation of low-lying energy levels and corresponding eigenvectors as well as the dynamic susceptibility $C(\omega)$ under equilibrium and non-equilibrium conditions as a function of interaction  $g$ or $\gamma$ and  disorder $ \sigma \hbar \bar \omega $ strengths. 

In the absence of interaction  the dynamic susceptibility $C(\omega)$ of disordered SQAs demonstrates $N$ small amplitude spikes transforming to sharp resonances as a small dissipation is included. These spikes have an origin in photon-induced transitions between various energy levels. An increase of interaction strength allowing to overcome the disorder, i.e. as $|g| >\sigma \hbar \bar \omega$, leads to an interesting regime in which all resonances except a single one are strongly suppressed. The amplitude of  dominant resonances $A_d(g)$ is strongly enhanced and increases with the number of qubits, $N$. We attributed the appearance of such dominant  resonance to the collective quantum-mechanical excited states occurring in disordered interacting SQAs. We obtain such collective behavior in various disordered SQAs with either short-range (Ising type) or artificially prepared global (exchange type) interactions.  

The dominant resonances with greatly enhanced amplitude indicating the presence of coherent collective quantum dynamics, have been observed in non-equilibrium dynamic susceptibility, $C_n(\omega)$ for disordered SQAs coupled to a low-dissipative resonator where photons are trapped in different Fock states, $|n>$ .The dominant resonances occurring in such a system indicate also the presence of an effective global interaction between all qubits that is due to emission and absorption of virtual cavity photons by well separated qubits.  In our numerical analysis we also observed the collective AC Stark effect with the characteristic shift of the dominant resonance frequency depending on the number of photons, $|n>$. The obtained effect can be used for resolving photon number states with a high accuracy using the typical transmission measurement setup. 
                                                          
\textbf{Acknowledgements}
We thank M. Lisitskiy, A. Ustinov and A. Zagoskin for insightful discussions. We acknowledge the financial support through the European Union’s Horizon 2020 research and innovation program under grant agreement No 863313.
\bibliography{general}

\end{document}